# Bloch-like waves in random-walk potentials based on supersymmetry

Sunkyu Yu, Xianji Piao, Jiho Hong, and Namkyoo Park*

*Photonic Systems Laboratory, School of EECS, Seoul National University, Seoul 151-744, Korea*

*E-mail address for correspondence: nkpark@snu.ac.kr*

**Bloch's theorem was a major milestone that established the principle of bandgaps in crystals. Although it was once believed that bandgaps could form only under conditions of periodicity and long-range correlations for Bloch's theorem, this restriction was disproven by the discoveries of amorphous media and quasicrystals. While network and liquid models have been suggested for the interpretation of Bloch-like waves in disordered media, these approaches based on searching for random networks with bandgaps have failed in the deterministic creation of bandgaps. Here, we reveal a deterministic pathway to bandgaps in random-walk potentials by applying the notion of supersymmetry to the wave equation. Inspired by isospectrality, we follow a methodology in contrast to previous methods: we transform order into disorder while preserving bandgaps. Our approach enables the formation of bandgaps in extremely-disordered potentials analogous to Brownian motion, and also allows the tuning of correlations while maintaining identical bandgaps, thereby creating a family of potentials with 'Bloch-like eigenstates'.**

# Introduction

The isospectral problem posed via the question "Can one hear the shape of a drum?"[1] introduced many fundamental issues regarding the nature of eigenvalues (sound) with respect to potentials (the shapes of drums). Following the demonstration presented in ref. 2, it was shown that it is not possible to hear the shape of a drum because of the existence of different drums (potentials) that produce identical sounds (eigenvalues), *a.k.a.* isospectral potentials. Although the isospectral problem has deepened our understanding of eigenstates with respect to potentials and raised similar questions in other physical domains[3], it has also resulted in various interesting applications such as the detection of quantum phases[4] and the modeling of anyons[5].

The field of supersymmetry[6] (SUSY) shares various characteristics with the isospectral problem. SUSY, which describes the relationship between bosons and fermions, has been treated as a promising postulate in theoretical particle physics that may complete the Standard Model[6]. Although the experimental demonstration of this postulate has encountered serious difficulties and controversy, the concept of SUSY and its basis of elegant mathematical relations have given rise to remarkable opportunities in many other fields, e.g., SUSY quantum mechanics[7] and topological modes[8]. Recently, techniques from SUSY quantum mechanics have been utilized in the field of optics, thereby enabling novel applications in phase matching and isospectral scattering[9-12], complex potentials with real spectra[13], and complex Talbot imaging[14].

In this paper, we propose a supersymmetric path for the generation of Bloch-like waves and bandgaps without the use of Bloch's theorem[15]. In contrast to approaches based on an iterative search for random networks[16-19] with bandgaps, a deterministic route toward bandgap creation in the case of disordered potentials is achieved based on the fundamental wave equation. This result not only demonstrates that long-range correlation is a sufficient but not a necessary condition for Bloch-like waves[16-19] but also enables the design of random-walk potentials with bandgaps. Such designs can facilitate the creation of a family of potentials with 'Bloch-like eigenstates': identical bandgaps and tunable long-range correlations, even extending to conditions of extreme disorder analogous to Brownian motion. We demonstrate that the counterintuitive phenomenon of 'strongly correlated wave behaviors in weakly correlated potentials' originates from the ordered modulation of potentials based on spatial information regarding the ground state, which is the nature of SUSY. We also show that our approach for Bloch-like waves can be extended to multi-dimensional potentials under a certain condition, allowing highly-anisotropic control of disorder.

## Results

**Relation between eigenstates and potential correlations**    To employ the supersymmetric technique[7,9], we investigate waves governed by the 1D Schrodinger-like equation, which is applicable to a particle in nonrelativistic quantum mechanics or to a transverse electric (TE) mode in optics. Without a loss of generality, we adopt conventional optics notations for the eigenvalue equation $H_o\psi = \gamma\psi$, where the Hamiltonian operator $H_o$ is

$$H_o = \frac{k^2}{k_0^2} + V_o(\mathbf{x}),\tag{1}$$

$k = i\partial_x$ is the wavevector operator, $k_0$ is the free-space wavevector, $V_o(x) = -[n(x)]^2$ is the optical potential, $n(x)$ is the refractive index profile, $\psi$ is the transverse field profile, and $n_{\text{eff}}$ is the effective modal index for the eigenvalue $\gamma = -n_{\text{eff}}^2$. Two independent methods are applied to Eq. (1) for verification, the Finite Difference Method[20] (FDM) and the Fourier Grid Hamiltonian[21] (FGH) method, whereby both yield identical results for the determination of bound states (see the Methods).

To examine the relationship between wave eigenspectra and the correlations of potentials, three types of random-walk potentials are analyzed: crystals, quasicrystals, and disordered potentials, which are generated by adjusting the refractive index profile. Figure 1a represents a 1D binary Fibonacci quasicrystal (the 6th generation with an inflation number, or sequence length, of $N = 8$, substituting

A→B and B→BA for each generation using A as the seed), where each element is defined by the gap between the high-index regions: A (or B) for a wider (or narrower) gap. The crystal and the disordered potential are generated using the same definition of elements, while the crystal has an alternating sequence (BABABA…), and the disordered potential has equal probabilities of A and B for each element (*i.e.,* it is a Bernoulli random sequence[22] with probability $p = 0.5$). To quantify the correlation, the Hurst exponent[23,24] $H$ is introduced (Fig. 1b; see the Methods). As $N$ increases, both the crystal and the quasicrystal have $H$ values that approach 0 (*i.e.,* they exhibit 'ballistic behavior' with strong negative correlations[24]), in stark contrast to the Bernoulli random potential, which has $H \sim 0.4$ (close to ideal Brownian motion, with $H = 0.5$).

Figures 1c-1h illustrate the stationary eigenstates for each potential, which are calculated using the FDM and the FGH method. Consistent with previous studies[16-19,25-29], Bloch-like waves with wide bandgaps are obtained for the ordered potentials of the crystal (Fig. 1c) and the quasicrystal (Fig. 1d), and the Bloch-like nature becomes more apparent with increasing $N$ (Figs 1f,1g). By contrast, no bandgap is observed for the Bernoulli random potential, which lacks *any* correlations (Fig. 1e), especially for larger $N$ (Fig. 1h); this lack of correlation originates from the broken coherence of this case, which hinders the destructive interference that is necessary for the formation of bandgaps. It should also be noted that many eigenstates are localized within this random potential, exhibiting a phenomenon that is widely known as Anderson localization[29].

**Supersymmetric transformation for quasi-isospectral design**  In light of the results in Fig. 1, we now consider the following question: "*Is it possible to design nearly uncorrelated (or Brownian) potentials with H ~ 0.5 while preserving the original bandgaps?*". To answer this question positively, we exploit the SUSY transformation to achieve quasi-isospectral potentials[7,9]. In Eq. (1), it is possible to decompose the Hamiltonian operator as follows: $H_o - \gamma_0 = NM$, where $N = -ik / k_0 + W(x)$, $M = ik / k_0 + W(x)$, $W(x)$ is the superpotential that satisfies the Riccati equation $W(x)^2 - i[kW(x)]/k_0 + \gamma_0 = V_o(x)$, and $\gamma_0$ is the ground-state eigenvalue of $H_o\psi_0 = \gamma_0\psi_0$. Then, the inversion of the $N$ and $M$ operators yields the SUSY Hamiltonian $H_s$ with the SUSY-partner potential $V_s(x)$: $H_s = MN + \gamma_0 = k^2 / k_0 + V_s(x)$. From the original equation $H_o\psi = \gamma\psi$, the relation $H_s \cdot (M\psi) = \gamma \cdot (M\psi)$ is obtained, thus proving isospectrality with $\gamma$ and the transformed eigenstates of $M\psi$[7,9]. For the later discussion of 2D potentials, it is noted that the isospectrality between $H_o$ and $H_s$ can also be expressed in terms of the intertwining relation $MH_o = H_sM$ = $MNM$ ($MH_o\psi = H_s(M\psi) = \gamma(M\psi)$ when $H_o\psi = \gamma\psi$), where the operator $M$ is the 'intertwining operator'[30,31].

The solution $W(x)$ is simply obtained from the Riccati equation through $W(x) = [\partial_x\psi_0(x)]/[k_0\psi_0(x)]$ for unbroken SUSY[7,9], which also provides the ground-state annihilation equation $M\psi_0 = [ik / k_0 + W(x)]\psi_0 = O$. Because $V_s(x) = -[n_s(x)]^2$ is equivalent to

$$V_s(x) = W^2 + i\frac{k}{k_0}W + \gamma_0, \tag{2}$$

the index profile $n_s(x)$ after the SUSY transformation can finally be obtained as follows:

$$[n_s(x)]^2 = [n(x)]^2 + 2\frac{1}{k_0{}^2}\frac{d}{dx}\left(\frac{\psi_0{}'}{\psi_0}\right) = [n(x)]^2 + 2\frac{1}{k_0{}^2}\frac{d^2}{dx^2}\left(\log\psi_0\right). \tag{3}$$

Equation (3) demonstrates that the SUSY transformation can be achieved deterministically based solely on the ground-state-dependent functionality. Figure 2 illustrates an example of serial SUSY transformations applied to the 1D Fibonacci quasicrystal potential defined in Fig. 1, where the small value of $N = 5$ is selected for clarity of presentation. For each SUSY transformation, all eigenstates of each previous potential, except for the ground state, are preserved in the transformed spatial profiles, while the shape of the designed potential becomes 'disordered' through 'deterministic' SUSY transformations.

**Potentials with Bloch-like states and tunable randomness** Because the presence of deterministic order is essential for Bloch-like waves and bandgaps, regardless of the presence of long-range correlations in their spatial profiles[16-19,25-29], the 'randomly shaped' potentials (Fig. 2) that can be 'deterministically' derived by applying SUSY transformations to ordered potentials offer the possibility of combining Bloch-like waves and disordered potentials. To investigate the wave behavior associated with the SUSY transformation, we consider a larger-$N$ regime in which the wave behaviors are clearly distinguished between ordered (Figs 1f,1g) and disordered potentials (Fig. 1h). Figures 3a and 3b present the results obtained after the $10^{th}$ SUSY transformation for the crystal (Figs 3a,3c) and the quasicrystal (Figs 3b,3d) with $N = 144$. Although the shapes of the SUSY-transformed potentials and the spatial information of the eigenstates in Figs 3a and 3b are markedly different from those of the corresponding original potentials in Figs 1c and 1d, the eigenspectrum of each potential is preserved, save for the annihilation of the 10 lowest eigenstates, which is consistent with the nature of SUSY transformations. From the SUSY transformation $M\psi = \{ik \ / \ k_0 + \partial_x\psi_0(x) \ / \ [k_0\psi_0(x)]\} \cdot \psi$, it is also expected that the distribution of the ground state $\psi_0(x)$ with respect to the original state $\psi$ primarily affects the effective width[32] of the transformed eigenstate $M\psi$. In crystals that have highly-overlapped intensity profiles between eigenstates, the effective width of $\psi$ decreases progressively from serial SUSY transformations due to the 'bound' distribution of $\psi_0(x)$. For a quasicrystal, the variation of the effective width showed more complex behavior owing to its spatially-separated eigenstates (see Supplementary Note 1 and Supplementary Figures 1, 2 for the comparison between crystal and quasicrystal potentials).

The eigenspectral conservation is apparent in Figs 3c and 3d, which depict the variation in the effective modal index that occurs during the SUSY transformations (up through the $20^{th}$ SUSY transformation). As shown, the eigenspectrum of each potential is maintained from the original to the $20^{th}$ SUSY transformation, save for a shift in the modal number, and therefore, the bandgaps in the remainder of the spectrum are maintained during the serial SUSY transformations (~125 states after the $20^{th}$ SUSY transformation following the loss of the 20 annihilated states). Consequently, bandgaps and Bloch-like eigenstates similar to those of the original potentials are allowed in SUSY-transformed potentials with disordered shapes (Figs 3a,3b) that can be classified as neither crystals nor quasicrystals.

Figures 4a–4h illustrate the shape evolutions of the crystal and quasicrystal potentials that are induced through the SUSY process, demonstrating the increase in disorder for both potentials. Again, note that the SUSY-based modulation is determined by Eq. (3), starting from the ground-state profile $\psi_0(x)$, which is typically concentrated near the center of the potential (Figs 3a,3b). To investigate the correlation features of SUSY-transformed potentials with bandgaps, we again consider the Hurst exponent. Figures 4i and 4j show the Hurst exponents for the transformed crystal and quasicrystal potentials as functions of the number of SUSY transformations for different sequence lengths ($N = 34$, 59, 85, and 144).

The figures show that, for successive applications of SUSY transformations, the Hurst exponents of the crystal and quasicrystal potentials ($H = 0 \sim 0.1$) increase and saturate at $H \sim 0.8$. For example, at $N = 144$, the negative correlations ($H < 0.5$) of the crystal and quasicrystal potentials ($H = 0 \sim 0.1$) become completely uncorrelated, with $H = 0.51$ after the $10^{th}$ SUSY transformation; the correlations are even weaker than that of the Bernoulli random potential ($H = 0.35 \sim 0.48$, Figs 1b,4i,4j) and approach the uncorrelated Brownian limit of $H = 0.5$. After the $10^{th}$ SUSY transformation, the correlation begins to increase again into the positive-correlation regime ($H \geq 0.5$, with long-lasting, *i.e.,* persistent, potential shapes), thereby exhibiting a transition between negative and positive correlations in the potentials. This transition from an 'anti-persistent' to 'persistent' shape originates from the smoothing of the original potential caused by the slowly varying term $\partial_x^2 \{log[\psi_0(x)]\}$ in Eq. (3), which is derived from the nodeless ground-state wavefunction $\psi_0(x)$.

We note that this $\psi_0(x)$-dependent modulation shows a dependence on the size (or sequence length $N$) of the potentials; for a potential with a large size, $\psi_0(x)$ varies weakly over a wide range, thus decreasing the relative strength of the SUSY-induced modulation $\partial_x^2 \{log[\psi_0(x)]\}$ (Figs 4i,4j). Thereby, the number of SUSY transformations required for extreme randomness ($H \sim 0.5$) increases with the size of the potential (Figs 4i,4j, ($S_B$, $N$) = (4, 34), (6, 55), (8, 89), and (10, 144), where $S_B$ is the required SUSY transformations for $H \sim 0.5$). Eventually, the SUSY transformation to periodic potentials of *infinite* size $n(x) = n(x + \Lambda)$ preserves the periodicity because the SUSY transformation with the Bloch ground state $\psi_0(x) = \psi_0(x + \Lambda)$ will repeatedly result in periodic potentials $n_s(x) = n_s(x + \Lambda)$.

These results reveal that the application of SUSY transformations to ordered (crystal or quasicrystal) potentials allows for remarkable control of the extent of the disorder while preserving Bloch-like waves and bandgaps. Therefore, a family of potentials with 'Bloch-like eigenstates', for its members have identical bandgaps but tunable disorders, can be constructed through the successive application of SUSY transformations to each ordered potential, with a range of disorder spanning almost the entire regime of Hurst exponents indicating negative and positive correlations ($0 \leq H \leq 0.8$), including the extremely uncorrelated Brownian limit of $H \sim 0.5$. As an extension, in Supplementary Note 2, we also provide the design strategy of random-walk discrete optical systems (composed of waveguides or resonators) that deliver Bloch-like bandgaps, starting from the $1^{st}$-order approximation of Maxwell's equations, i.e., coupled mode theory[33,34].

**Extension of SUSY transformations to 2D potentials**    In stark contrast to the case of 1D potentials, which exclusively satisfy a 1:1 correspondence between their shape and ground state[7], it is more challenging to achieve isospectrality in multi-dimensional potentials. Although studies have shown the vector-form SUSY decomposition of multi-dimensional Hamiltonians[35-37], such an approach, which is analogous to the Moutard transformation[38], cannot guarantee isospectrality. This approach only generates a pair of scalar Hamiltonians with eigenspectra that, in general, do not overlap but together compose the eigenspectrum of the other vector-form Hamiltonian[35-37]. Here, we employ an alternative route[30,31,39] starting from the intertwining relation $MH_o = H_sM$ to implement a class of multi-dimensional isospectral potentials.

Without the loss of generality, we consider the 2D Schrodinger-like equations with the Hamiltonian of $H_o = -(1/k_0^2) \cdot \nabla^2 + V_o(x,y)$ and its SUSY-partner Hamiltonian $H_s = -(1/k_0^2) \cdot \nabla^2 + V_s(x,y)$. To satisfy the intertwining relation $MH_o = H_sM$, the ansatz for the intertwining operator $M$ can be introduced[30,31], similarly to the 1D case:

$$M = M_o + M_d = M_o(x,y) + M_x(x,y) \cdot \frac{\partial}{\partial x} + M_y(x,y) \cdot \frac{\partial}{\partial y}, \qquad (4)$$

where $M_o$, $M_x$, and $M_y$ are arbitrary functions of $x$ and $y$. From Eq. (4), the intertwining relation $MH_o = H_sM$ can be expressed in terms of operator commutators as follows:

$$[\nabla^2, M_d] = -[\nabla^2, M_o] + k_0^2 \cdot ([V_o, M_d] + V_dM),$$ (5)

where $V_d$ is the modification of the potential through the SUSY transformation: $V_d(x,y) = V_s(x,y) - V_o(x,y)$. Although here we focus on the 2D example, it is noted that Eq. (4) can be generalized to $N$-dimensional problems[30,31] as $M = M_o(x_1, x_2, \ldots, x_N) + \sum M_i(x_1, x_2, \ldots, x_N) \cdot \partial_i$ while maintaining Eq. (5).

The derivation in the Methods (Eqs (6-21)) starting from Eq. (5) demonstrates that the procedure of the 1D SUSY transformation can be applied to a 2D potential for each $x$- and $y$-axis independently, when the potential satisfies the condition of $V_o(x,y) = V_{ox}(x) + V_{oy}(y)$. We also note that serial 2D SUSY transformations are possible because the form of $V_o(x,y) = V_{ox}(x) + V_{oy}(y)$ is preserved during the transformation, consequently deriving a family of 2D quasi-isospectral potentials. Figure 5 shows an example of SUSY transformations in 2D potentials, maintaining Bloch-like eigenstates. Both the $x$- and $y$-axis cross-sections of the 2D original potential $V_o(x,y) = V_{ox}(x) + V_{oy}(y)$ have profiles of $N = 8$ binary sequences (Fig. 5a), as defined in Fig. 1. Following the procedure of Eqs (17-21) in the Methods, we apply SUSY transformations to the $x$- and $y$-axes separately, achieving the highly anisotropic shape of the potential as shown in Fig. 5b (the $5^{th}$ $x$-axis SUSY transformed potential) and Fig. 5c (the $5^{th}$ $y$-axis SUSY transformed potential). It is evident that this anisotropy can be controlled by changing the number of SUSY transformations for the $x$- and $y$-axes independently, and the isotropic application of SUSY transformations recovers the isotropic potential shape (Fig. 5d). Regardless of the number of SUSY transformations and their anisotropic implementations, the region of bandgaps of the original potential is always preserved (Fig. 5e). Interestingly, the annihilation by 2D SUSY transformation occurs not only in the ground state but also in all of the excited states sharing a common 1D ground-state profile (for details see the Methods, Supplementary Note 3 and Supplementary Figure 9). Consequently, the width of the bandgap can be slightly changed owing to the annihilation of some excited states near the bandgap.

To investigate the correlation features of 2D SUSY-transformed potentials, we quantify the angle-dependent degree of the correlation. Figures 5f and 5g show the angle-dependent variation of the Hurst exponent for the anisotropic (the $5^{th}$ $x$-axis SUSY-transformed potential, Fig. 5b) and isotropic (the $5^{th}$ $x$- and $y$-axis SUSY transformed potential, Fig. 5d) disordered potential. Compared to the original potential (gray symbols in Figs 5f,5g, for Fig. 5a), $H$ increases along the axis with the SUSY transformations ($x$-axis in Fig. 5f and $x$- and $y$-axes in Fig. 5g). The potential is disordered at all angles, especially in the diagonal directions ($\pm 45°$), owing to the projection of the SUSY-induced disordered potential shapes (45° profiles in Figs 5b-5d).

## Discussion

To summarize, by employing supersymmetric transformations, we revealed a new path toward the deterministic creation of random-walk potentials with 'crystal-like' wave behaviors and tunable spatial correlations, extending the frontier of disorder for Bloch-like waves and identical bandgaps. Despite their weak correlations and disordered shapes, SUSY-transformed potentials retain the deterministic 'eigenstate-dependent order' that is the origin of bandgaps, which is in contrast to the hyperuniform[18,40-42] disorder of pointwise networks and deterministic aperiodic structures such as quasicrystals[28,43] or the Thue-Morse[44] and Rudin-Shapiro[45] sequences. We also extend our discussion to multi-dimensions, achieving highly-anisotropic or quasi-isotropic disordered 2D potentials, while preserving bandgaps. Our results, which were obtained based on a Schrodinger-like equation, reveal a novel class of Bloch-wave disorder that approaches the theoretical limit of Brownian motion while maintaining wide bandgaps identical to those of existing crystals or quasicrystals in both electronics and optics. We further envisage a novel supersymmetric relation, based on the famous SUSY theory in particle physics, between ordered potentials and disordered potentials with coherent wave behaviors in solid-state physics. The extension of the SUSY transformation to non-Schrodinger equations, $e.g.$, transverse magnetic modes in electromagnetics (as investigated in Supplementary of ref. 11), or to the approximated Hamiltonians applicable to arbitrary-polarized optical elements (Supplementary Note 2) will be of importance for future applications, *e.g.*, polarization-independent bandgaps based on dual-polarized eigenstates[46].

**Methods**

**Details of the FDM and FGH method**    The FDM utilizes the approximation of the 2nd-derivative operator in the discrete form[20], and the FGH method, as a spectral method, uses a planewave basis with operator-based expressions in a spatial domain[21]. In both methods, the Hamiltonian matrices are Hermitian because of the real-valued potentials, thus enabling the use of Cholesky decomposition to solve the eigenvalue problem. To ensure an accurate SUSY process, Rayleigh quotient iteration is also applied to obtain the ground-state wavefunction. The boundary effect is minimized through the use of a buffer region ($n = 1.5$) of sufficient length ($30\ \mu m = 20\lambda_0$) on each side. Deep-subwavelength grids ($\Delta = 20$ nm $= \lambda_0/75$) are also used for the discretization of both the 1D and 2D potentials.

**Calculation of the Hurst exponent**    First, the discretized refractive index $n_p$ ($p = 1,2,\ldots,N$) is obtained at $x_p = x_{\text{left}} + (p\text{-}1)\cdot\Delta$, where $x_{\text{left}}$ is the left boundary of the potential, which is of length $L = (N\text{-}1)\cdot\Delta$. Partial sequences $X_q$ of $n_p$ for different length scales $d$ are then defined ($2 \leq d \leq N$ and $1 \leq q \leq d$). For the mean-adjusted sequence $Y_q = X_q - m$, where $m$ is the mean of $X_q$, we define the cumulative deviate series $Z_r$ as

$$Z_r = \sum_{q=1}^{r} Y_q \,. \tag{6}$$

The range of cumulative deviation is defined as $R(d) = max(Z_1,Z_2,\ldots,Z_d) - min(Z_1,Z_2,\ldots,Z_d)$. Using the standard deviation $S(d)$ of $Y_q$, we can now apply the power law to the rescaled range $R(d)/S(d)$ as follows:

$$E\left[\frac{R(d)}{S(d)}\right] = c_0 d^H \,. \tag{7}$$

This yields $log(E[R(d)/S(d)]) = H\cdot log(d) + c_1$, where $E$ is the expectation value and $c_0$ and $c_1$ are constants. $H$ is then obtained through linear polynomial fitting: $H = 0.5$ for Brownian motion, $0 \leq H <$ 0.5 for long-term negative correlations with switching behaviors, and $0.5 < H \leq 1$ for long-term positive correlations such that the sign of the signal is persistent.

**The condition for 2D isospectral potentials**   By assigning $M = M_o + M_d$ (Eq. (4)) to the intertwining relation $MH_o = H_sM$ with explicit forms of $H_o$ and $H_s$, it becomes $(M_o + M_d) \cdot [-(1/k_0^2) \cdot \nabla^2 + V_o] = [-(1/k_0^2) \cdot \nabla^2 + V_s] \cdot (M_o + M_d)$. Thus, we obtain Eq. (5); $[\nabla^2, M_d] = -[\nabla^2, M_o] + k_0^2 \cdot ([V_o, M_d] + V_dM)$, where $V_d = V_s - V_o$. Each commutator in Eq. (5) is also expressed as

$$
\begin{aligned}
[\nabla^2, M_d] &= (\nabla^2 M_x) \cdot \frac{\partial}{\partial x} + (\nabla^2 M_y) \cdot \frac{\partial}{\partial y} \\
&+ 2 \cdot \left[ \left( \frac{\partial M_x}{\partial x} \right) \cdot \frac{\partial^2}{\partial x^2} + \left( \frac{\partial M_y}{\partial y} \right) \cdot \frac{\partial^2}{\partial y^2} \right] + 2 \cdot \left( \frac{\partial M_y}{\partial x} + \frac{\partial M_x}{\partial y} \right) \cdot \frac{\partial^2}{\partial x \partial y}
\end{aligned}
\tag{8}
$$

$$
[\nabla^2, M_o] = (\nabla^2 M_o) + 2 \cdot \left[ \left( \frac{\partial M_o}{\partial x} \right) \cdot \frac{\partial}{\partial x} + \left( \frac{\partial M_o}{\partial y} \right) \cdot \frac{\partial}{\partial y} \right],
\tag{9}
$$

$$
[V_o, M_d] = -\left[ M_x \cdot \left( \frac{\partial V_o}{\partial x} \right) + M_y \cdot \left( \frac{\partial V_o}{\partial y} \right) \right].
\tag{10}
$$

It is noted that the higher-order ($\geq 2$) derivatives in Eq. (5) originate from the third and fourth terms in the RHS of Eq. (8). Comparing Eq. (8) with Eqs (9,10), all of the higher-order derivatives should be removed to satisfy Eq. (5). This then directly leads to the preconditions $M_x$ and $M_y$; $\partial_x M_x = 0$, $\partial_y M_y = 0$, and $\partial_x M_y + \partial_y M_x = 0$ which hold only for $M_x(x,y) = M_x(y) = a_x - by$ and $M_y(x,y) = M_y(x) = a_y + bx$ where $a_x, a_y$ and $b$ are arbitrary constants.

By applying $M_x(y)$, $M_y(x)$, and Eqs (8-10) to Eq. (5), we achieve two linear and one nonlinear equations for three unknowns $M_o$, $V_o$, and $V_d$ as

$$
\frac{\partial M_o}{\partial x} = \frac{1}{2} \cdot k_0^2 V_d \cdot (a_x - by),
\tag{11}
$$

$$
\frac{\partial M_o}{\partial y} = \frac{1}{2} \cdot k_0^2 V_d \cdot (a_y + bx),
\tag{12}
$$

$$(-\nabla^2 + k_0{}^2 V_d) M_o = k_0{}^2 \left[ (a_x - by) \cdot \left( \frac{\partial V_o}{\partial x} \right) + (a_y + bx) \cdot \left( \frac{\partial V_o}{\partial y} \right) \right]. \tag{13}$$

As a particular solution, we consider the case of $b = 0$ for simplicity. In this case, from Eqs (11,12), $M_o$ and $V_d$ are determined in the form of $M_o = f(\rho)$ and $V_d = \partial_\rho f(\rho)$, where $\rho = k_0{}^2 \cdot (a_x x + a_y y) / 2$ is the transformed coordinate and $f$ is an arbitrary function of $\rho$. By substituting $M_o$ and $V_d$, Eq. (13) then becomes

$$\frac{\partial V_o}{\partial \rho} = \frac{2}{k_0{}^2 (a_x{}^2 + a_y{}^2)} \cdot f(\rho) \cdot \frac{\partial f}{\partial \rho} - \frac{1}{2} \cdot \frac{\partial^2 f}{\partial \rho^2}, \tag{14}$$

which reveals the proper form of the 2D potential $V_o$ for SUSY transformations

$$V_o(\rho, \xi) = \frac{1}{k_0{}^2 (a_x{}^2 + a_y{}^2)} \cdot f^2(\rho) - \frac{1}{2} \cdot \frac{\partial f}{\partial \rho} + g(\xi), \tag{15}$$

where $\xi = k_0{}^2 \cdot (c_x x + c_y y) / 2$ is the transformed coordinate perpendicular to $\rho$, with $a_x \cdot c_x + a_y \cdot c_y = 0$. The supersymmetric potential $V_s = V_o + V_d$ then becomes

$$V_s(\rho, \xi) = \frac{1}{k_0{}^2 (a_x{}^2 + a_y{}^2)} \cdot f^2(\rho) + \frac{1}{2} \cdot \frac{\partial f}{\partial \rho} + g(\xi). \tag{16}$$

Using Eqs (15,16), now we can implement the procedure of serial 2D SUSY transformations. First, because of Eq. (15), $V_o$ should have the form of $V_o(\rho, \xi) = V_{o\rho}(\rho) + V_{o\xi}(\xi)$ for two Cartesian axes of $\rho$ and $\xi$. In this case, the corresponding $f(\rho)$ is obtained by solving the following Riccati equation:

$$V_{o\rho}(\rho) = \frac{1}{k_0{}^2 (a_x{}^2 + a_y{}^2)} \cdot f^2(\rho) - \frac{1}{2} \cdot \frac{\partial f}{\partial \rho} + \gamma_{o\rho}; \tag{17}$$

Its particular solution is listed as $f(\rho) = -[k_0{}^2 \cdot (a_x{}^2 + a_y{}^2) \cdot \partial_\rho \varphi_0(\rho)] / [2 \cdot \varphi_0(\rho)]$ (ref. 47); where $\varphi_0(\rho)$ is the nodeless ground state with the eigenvalue $\gamma_{o\rho}$ in the corresponding 1D Schrodinger-like equation

$$-\frac{k_0{}^2}{4} (a_x{}^2 + a_y{}^2) \cdot \frac{\partial^2 \varphi}{\partial \rho^2} + V_{o\rho}(\rho)\varphi = \gamma_\rho \varphi. \tag{18}$$

With the obtained $f(\rho)$, we finally achieve the SUSY-transformed potential along the $\rho$-axis satisfying the isospectrality, $V_s(\rho,\xi) = V_o(\rho,\xi) + \partial_\rho f(\rho)$, or

$$V_s(\rho,\xi) = V_o(\rho,\xi) - \frac{1}{2} \cdot k_0^2 (a_x^2 + a_y^2) \cdot \frac{d^2}{d\rho^2}(\log \varphi_o(\rho)). \tag{19}$$

Equivalently, the SUSY-transformation along the $\xi$ axis is

$$V_s(\rho,\xi) = V_o(\rho,\xi) - \frac{1}{2} \cdot k_0^2 (c_x^2 + c_y^2) \cdot \frac{d^2}{d\xi^2}(\log \phi_o(\xi)), \tag{20}$$

where $\phi_0(\xi)$ is the nodeless ground state with the eigenvalue $\gamma_{o\xi}$ in the following equation:

$$-\frac{k_0^2}{4}(c_x^2 + c_y^2) \cdot \frac{\partial^2 \phi}{\partial \xi^2} + V_{o\xi}(\xi)\phi = \gamma_\xi \phi. \tag{21}$$

Note that, after the SUSY transformation for the $\rho$- or $\xi$-axis, $V_s(\rho,\xi)$ still preserves the form of $V_s(\rho,\xi) = V_{s\rho}(\rho) + V_{s\xi}(\xi)$ (Eqs (19,20)), which is the necessary condition for the SUSY transformation of 2D potentials. Therefore, serial SUSY transformations can be applied to 2D arbitrary potentials of the form $V_o(\rho,\xi) = V_{o\rho}(\rho) + V_{o\xi}(\xi)$, and the level of SUSY transformations can be controlled independently for each axis, allowing highly anisotropic potential profiles. In addition, by assigning nonzero $b$, the allowed potential of $V_o(x,y)$ can be extended to non-separated forms[30,31].

**The eigenstate annihilation in 2D SUSY transformations**  In stark contrast to the ground-state annihilation in 1D SUSY transformations, the annihilation by 2D SUSY transformations is not restricted to the ground state. For simplicity, consider the case of $a_x = 2/k_0^2$ and $a_y = 0$ for $\rho = x$ and $\xi = y$ without any loss of generality. The Hamiltonian, which can be SUSY-transformed, is then expressed as $H_o = -(1/k_0^2)\cdot\nabla^2 + V_{ox}(x) + V_{oy}(y)$ for the eigenvalue equation $H_o\psi = \gamma\psi$. For the $x$-axis SUSY-transformation, the following equation should be satisfied to annihilate the SUSY-transformed eigenstate because $M\psi = O$:

$$-\frac{2}{k_0{}^2} \cdot \left(\frac{1}{\varphi_o} \cdot \frac{d\varphi_o(x)}{dx} - \frac{\partial}{\partial x}\right)\psi(x, y) = 0 \,. \tag{22}$$

Note that Eq. (22) is satisfied when $\psi(x,y) = \varphi_0(x) \cdot \phi(y)$, allowing the separation of variables in the 2D eigenvalue equation $H_o\psi = \gamma\psi$ as

$$\left[-\frac{1}{k_0{}^2} \cdot \frac{1}{\varphi_o} \cdot \frac{d^2\varphi_o(x)}{dx^2} + V_{ox}(x)\right] + \left[-\frac{1}{k_0{}^2} \cdot \frac{1}{\phi} \cdot \frac{d^2\phi(y)}{dy^2} + V_{oy}(y)\right] = \gamma \,. \tag{23}$$

It is noted that the first brace has the fixed constant of $\gamma_{ox}$, the *ground-state eigenvalue* of the 1D Schrodinger-like equation with the potential $V_{ox}(x)$. Meanwhile, because the second brace can be *any eigenvalues* of the solution $\phi(y)$ in the 1D Schrodinger-like equation with the potential $V_{oy}(y)$, it is clear that the annihilation by 2D SUSY transformation occurs not only in the ground state but also in all of the excited states sharing $\varphi_0(x)$. The detailed illustration of this result is shown in Supplementary Note 3 and Supplementary Figure 9.

**Acknowledgments**

The authors would like to thank M. A. Miri for their discussions on supersymmetric optics, especially with regard to ground-state annihilation and S. Torquato for the encouragement of our results and the introduction of hyperuniformity. This work was supported by the National Research Foundation of Korea through the Global Frontier Program (GFP) NRF-2014M3A6B3063708, the Global Research Laboratory (GRL) Program K20815000003, and the Brain Korea 21 Plus Project in 2015, which are all funded by the Ministry of Science, ICT & Future Planning of the Korean government.

**Author Contributions**

S.Y. and N.P. conceived of the presented idea. S.Y. developed the theory and performed the computations. X.P. and J.H. verified the analytical methods. N.P. encouraged S.Y. to investigate supersymmetry and supervised the findings of this work. All authors discussed the results and contributed to the final manuscript.

**Competing Interests Statement**

The authors declare that they have no competing financial interests.

**Figure Legends**

**Figure 1. Relation between eigenstates and potential correlations** **a**, Definitions of elements, illustrated for an example of a 1D Fibonacci quasicrystal ($N = 8$). $g_A = 600$ nm, $g_B = 200$ nm, $w = 120$ nm, $s = 140$ nm, and the wavelength is $\lambda_0 = 1500$ nm. **b**, Hurst exponent $H$ for each potential as a function of the sequence length $N$. The sequence lengths $N$ are selected to be equal to those of Fibonacci quasicrystals. The $H$ of the Bernoulli random potential is plotted with standard deviation error bars for 200 statistical ensembles. The black dashed line represents the Hurst exponent of ideal Brownian motion ($H = 0.5$). **c-e**, Eigenstates of each potential. The blue curve represents the ground state of each potential, and the colored lines represent the spectral ($n_{eff}$) distributions of the eigenstates. **f-h**, Evolutions of the band structures for different sequence lengths $N$: **c,f**, for crystals, **d,g**, for quasicrystals, and **e,h**, for Bernoulli random potentials. Note that the eigenstate inside the gap in **c,f**, is a surface state for an even $N$ (or an odd number of high-index regions) from the finite sizes of the potentials.

**Figure 2. SUSY transformation for quasi-isospectral potentials** A 1D Fibonacci quasicrystal ($N = 5$) is considered as an example. **a**, Original potential. **b-f**, 1st - 5th SUSY-transformed potentials. The orange (or black) dotted lines represent the preserved (or annihilated) eigenstates. All eigenstates are calculated using both the FDM and FGH method, the results of which are in perfect agreement.

**Figure 3. Eigenstates of SUSY-transformed crystals and quasicrystals** The 10th SUSY-transformed potentials and their eigenstates are depicted for **a**, a crystal potential and **b**, a quasicrystal potential. **c** and **d** show the eigenvalues of the SUSY-transformed potentials as a function of the modal numbers of the crystal and quasicrystal potentials, respectively. The 0th SUSY-transformed potential corresponds to the original potential. $N = 144$.

**Figure 4. Correlation features of SUSY-transformed potentials** The evolutions of the potential profiles following the successive application of SUSY transformations ($0^{th}$, $6^{th}$, $12^{th}$, and $18^{th}$) for **a,** a crystal and **e,** a quasicrystal. **b-d** and **f-h** present magnified views (at $x_L$, $x_C$, and $x_R$) of the potentials for even numbers of SUSY transformations (overlapped, up to the $20^{th}$ transformation; the blue arrows indicate the direction of potential modulation). $N = 144$ in **a-h**. Hurst exponents $H$ as functions of the number of SUSY transformations are shown for **i,** crystals and **j,** quasicrystal potentials, with different sequence lengths ($N = 34$, $59$, $85$, and $144$). The red (or blue) region represents the regime of positive (or negative) correlation, whereas the white region corresponds to the uncorrelated Brownian limit. The arrow indicates the regime of the Bernoulli random potential (Fig. 1b).

**Figure 5. 2D SUSY-transformed potentials with bandgaps maintained** The evolutions of the potential profiles following the application of SUSY transformations to the $x$- and $y$-axes are shown: **a,** original, **b,** $x$-axis SUSY transformed ($5^{th}$). **c,** $y$-axis SUSY transformed ($5^{th}$), and **d,** $x$- and $y$-symmetric SUSY-transformed ($x$-axis: $5^{th}$, $y$-axis: $5^{th}$) potentials. The spatial profiles of the potentials for $0°$ ($x$-axis), $45°$, and $90°$ ($y$-axis) are also overlaid in **a-d**. **e** shows the eigenvalues of the SUSY-transformed potentials as a function of the modal numbers. The gray regions denote bandgaps. The total SUSY number is the sum of the numbers of SUSY transformations for $x$- and $y$-axes (*i.e.*, 5 for both **b** and **c**, and 10 for **d**). **f** and **g** are the Hurst exponents for different directions of the 2D potentials that are **f,** highly anisotropic ($x$-axis: $5^{th}$, $y$-axis: $0^{th}$) and **g,** quasi-isotropic ($x$-axis: $5^{th}$, $y$-axis: $5^{th}$) SUSY transformations. The gray symbols in **f** and **g** are the Hurst exponents of the original potential.

# Figures

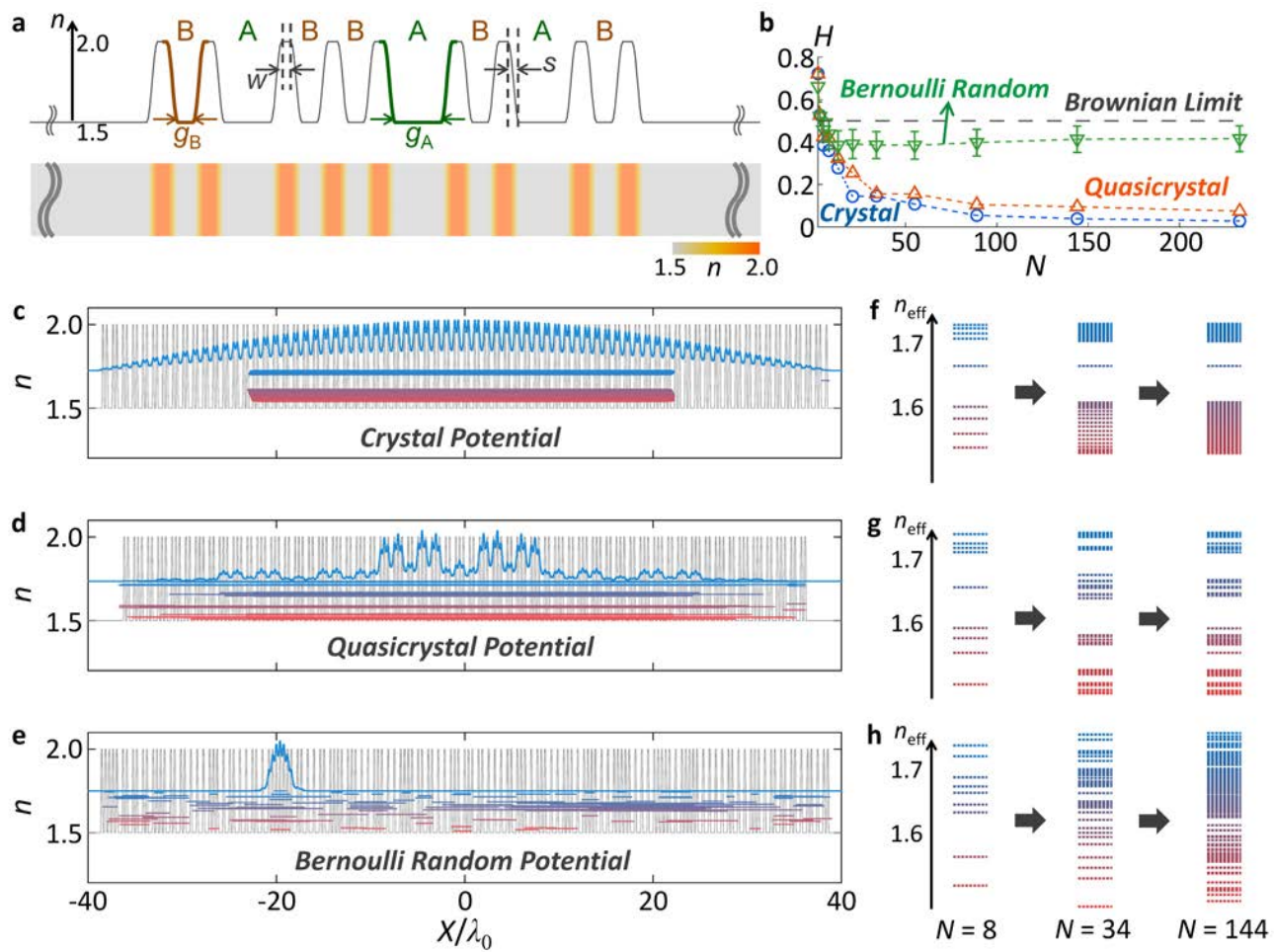

**Yu, Figure 1**

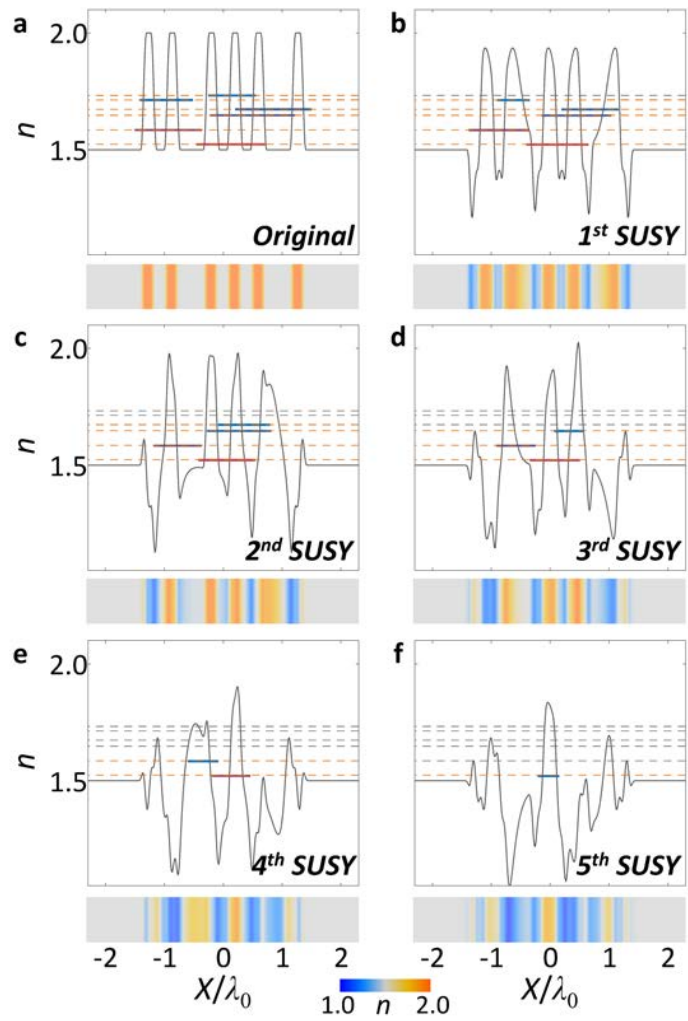

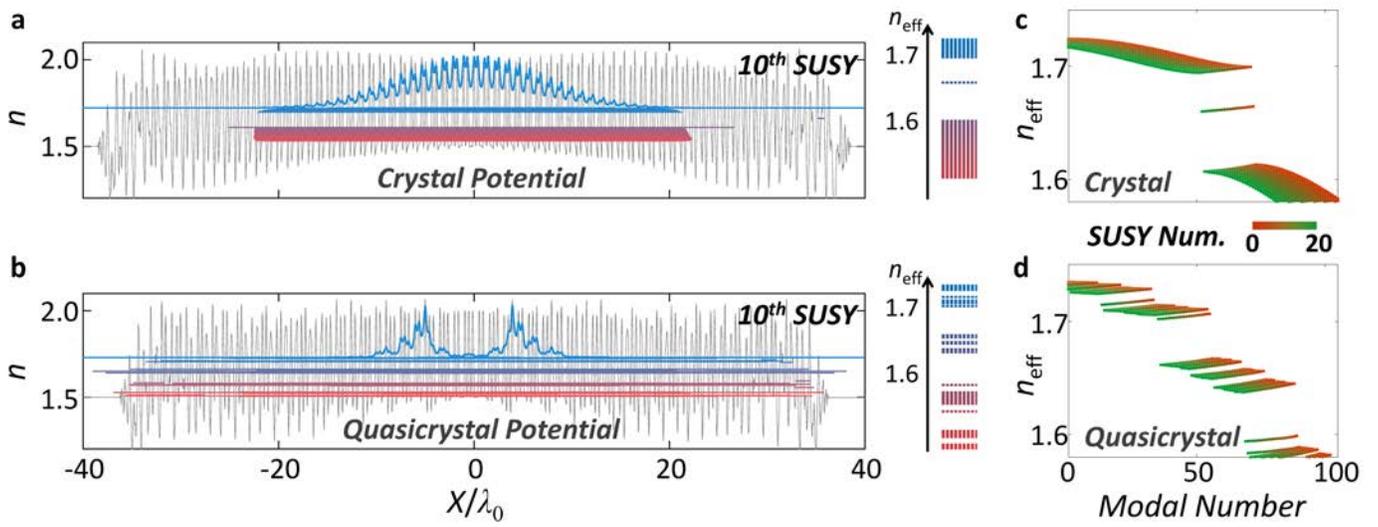

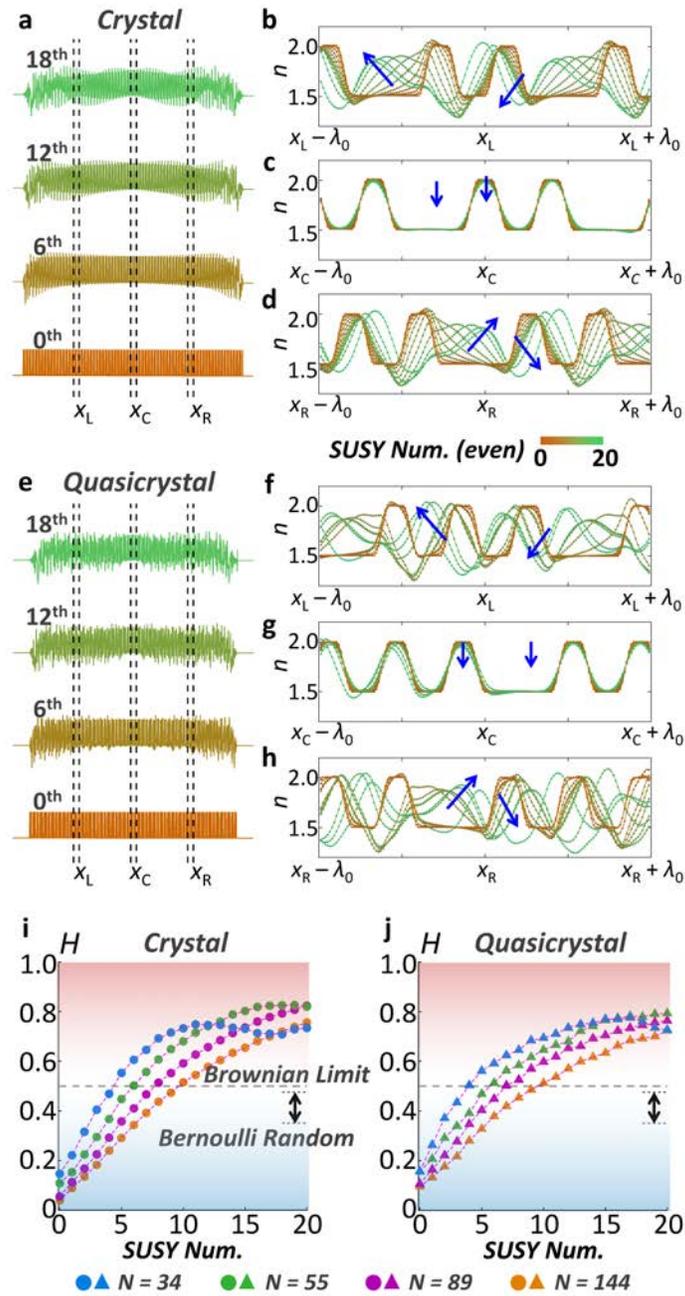

**a** *Crystal*

**b**

**c**

**d**

*SUSY Num. (even)* 0 — 20

**e** *Quasicrystal*

**f**

**g**

**h**

**i** *H* *Crystal*

**j** *H* *Quasicrystal*

● ▲ *N = 34*  ● ▲ *N = 55*  ● ▲ *N = 89*  ● ▲ *N = 144*

**Yu, Figure 4**

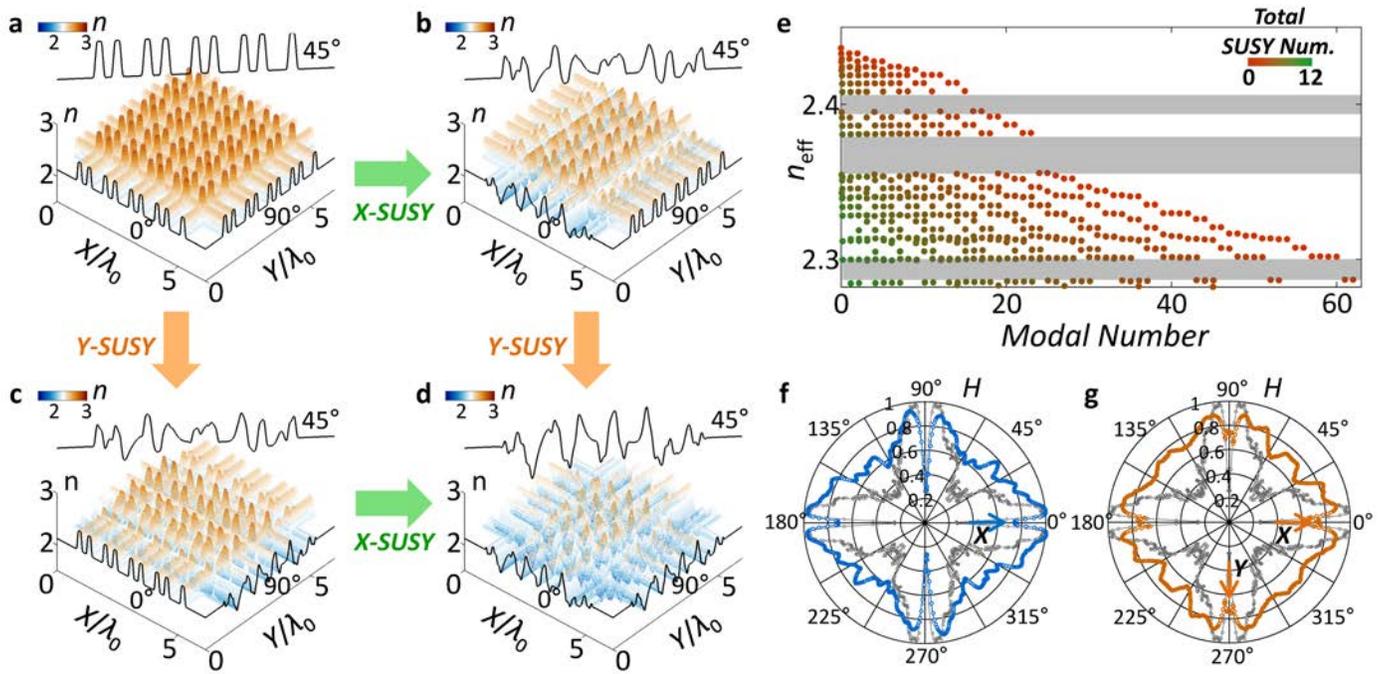

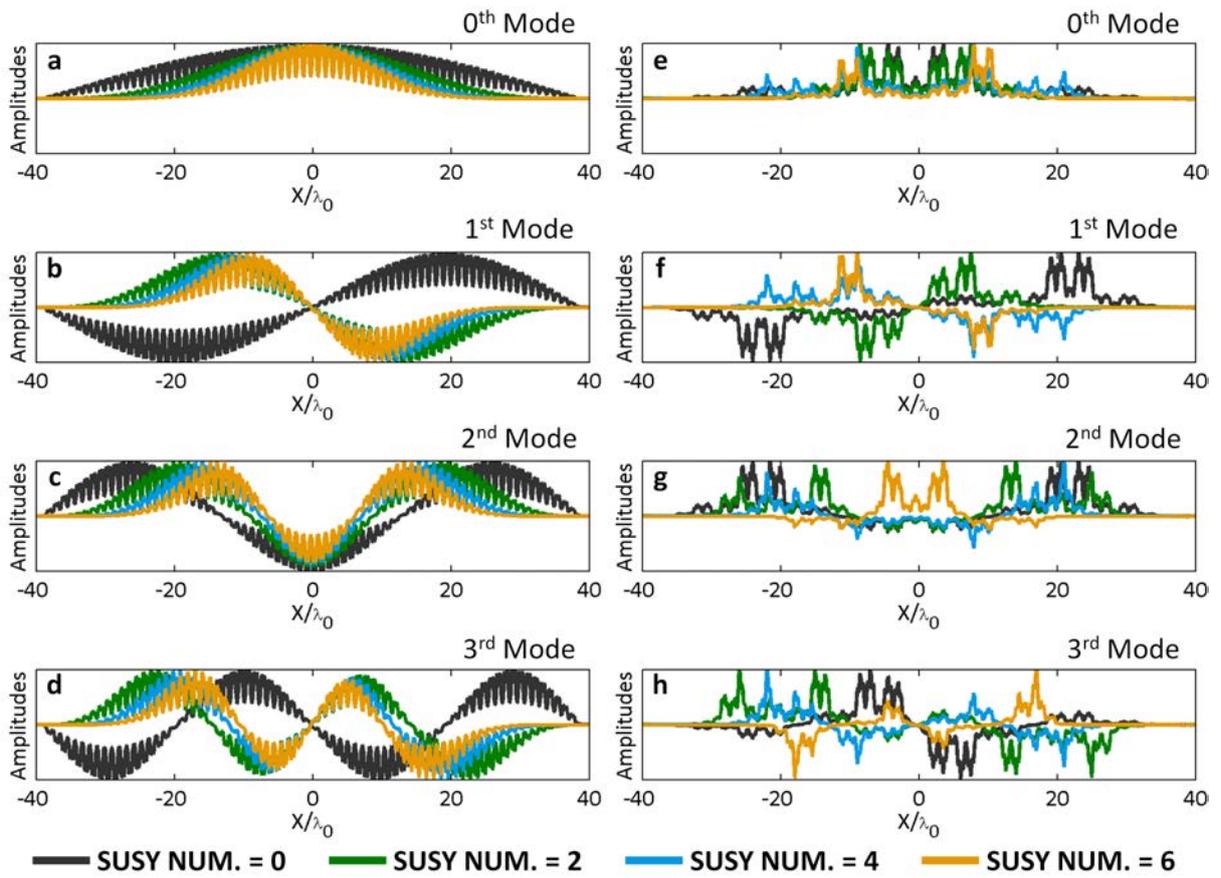

**Supplementary Figure 1. The spatial variation of eigenstates from SUSY transformations** (black: 0, green: 2, blue: 4, orange: 6 SUSY transformations): crystals (**a-d**) and quasicrystals (**e-f**) for different mode numbers (**a,e,**: $0^{th}$, **b,f,**: $1^{st}$, **c,g,**: $2^{nd}$, **d,h,**: $3^{rd}$ mode). $N = 144$ and the initial shapes of potentials are same as those used in the main manuscript.

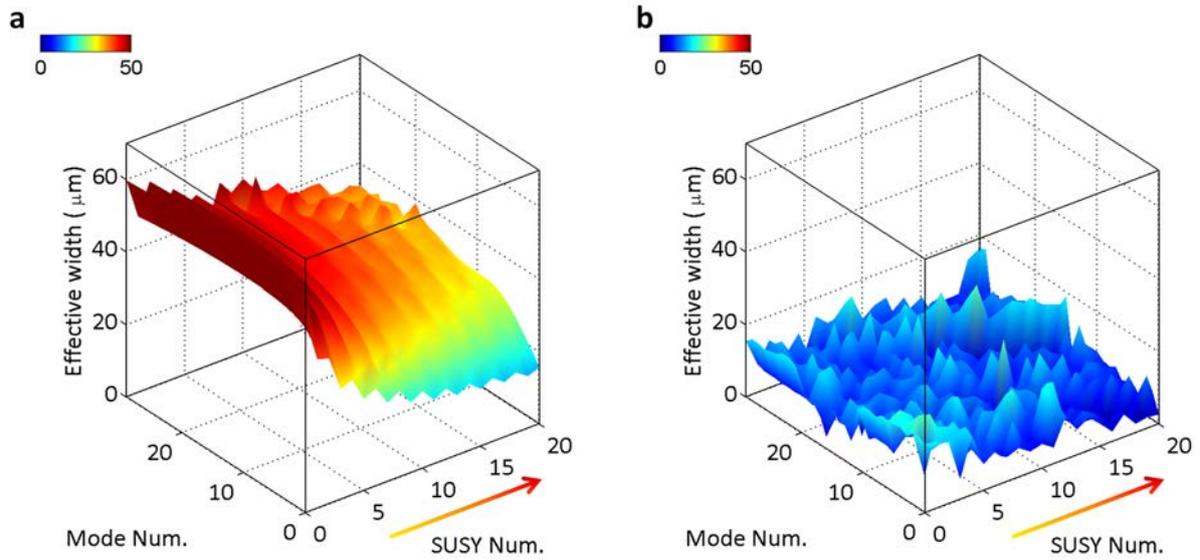

**Supplementary Figure 2. The variation of the effective widths of the eigenstates from SUSY transformations: a,** in the crystal, and **b,** in the quasicrystal. The number of SUSY transformations changes from 0 to 20. $N = 144$ and the initial shapes of potentials are same as those used in the main manuscript.

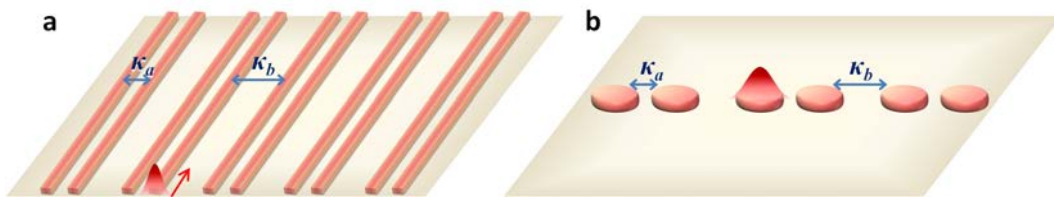

**Supplementary Figure 3. Binary photonic molecules a,** A waveguide-based photonic molecule of the spatial CMT ($\xi = x$). **b,** A resonator-based photonic molecule of the temporal CMT ($\xi = t$).

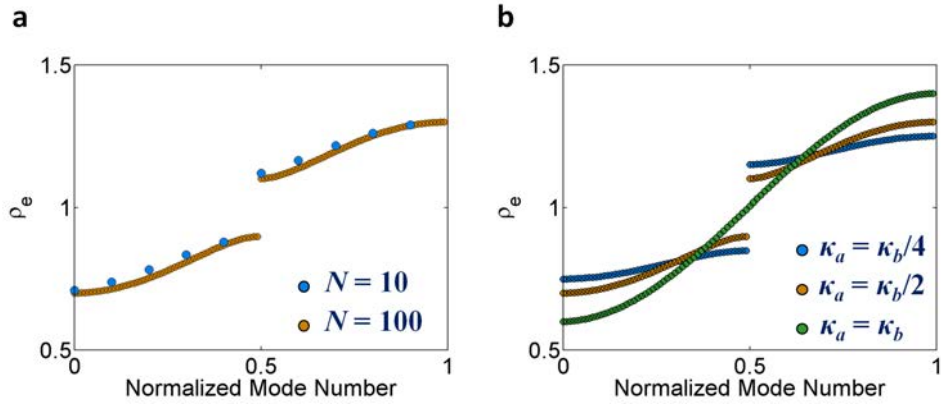

**Supplementary Figure 4. Eigenspectra of binary photonic molecules** for **a,** different $N$ ($\kappa_a$ = 0.1) and **b,** different $\kappa_a$ ($N$ = 100). $\rho_0$ = 1 and $\kappa_b$ = 0.2 for all cases.

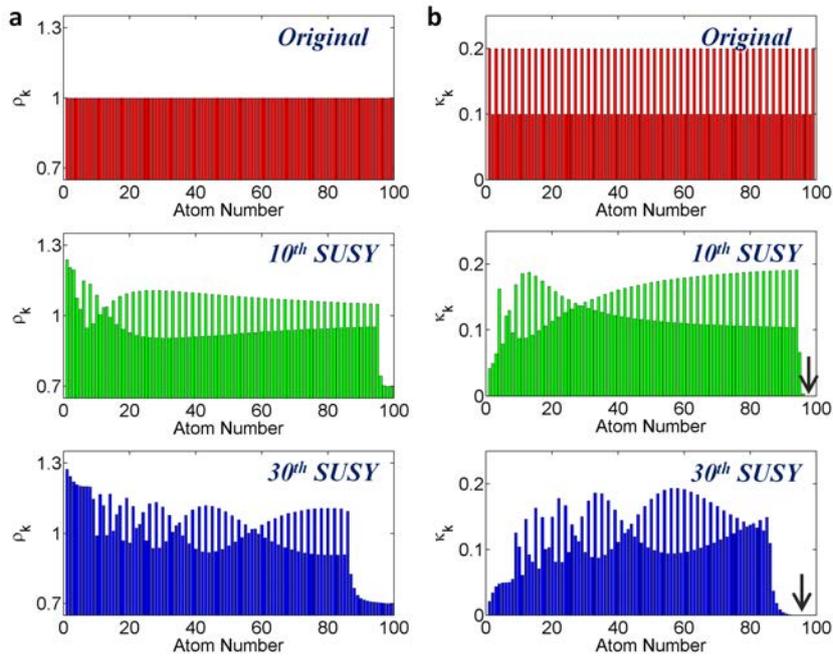

**Supplementary Figure 5. The variation of the CMT parameters from the SUSY transformations a,** The self-evolution for each atom and **b,** the coupling between atoms for the original, $10^{th}$, and $30^{th}$ SUSY-transformed photonic molecules. The original binary molecule has the following parameters: $\kappa_a$ = 0.1, $\kappa_b$ = 0.2, $\rho_0$ = 1 and $N$ = 100. Black arrows denote the decoupling.

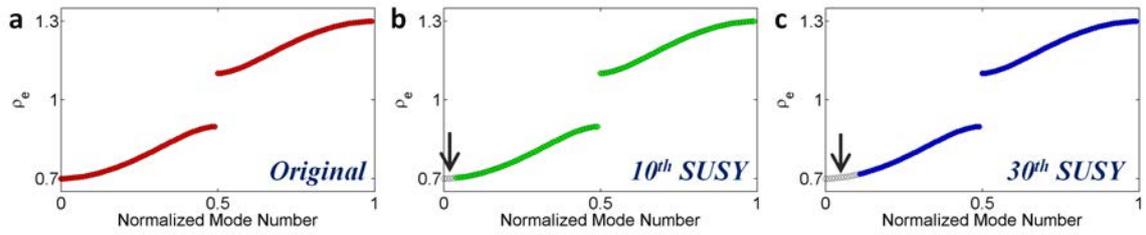

**Supplementary Figure 6. Eigenspectra of SUSY-transformed photonic molecules in Supplementary Fig. 3 a,** Original binary photonic molecule. **b,** The 10[th] and **c,** 30[th] SUSY-transformed photonic molecules. Black arrows denote the eigenstates of the decoupled atoms.

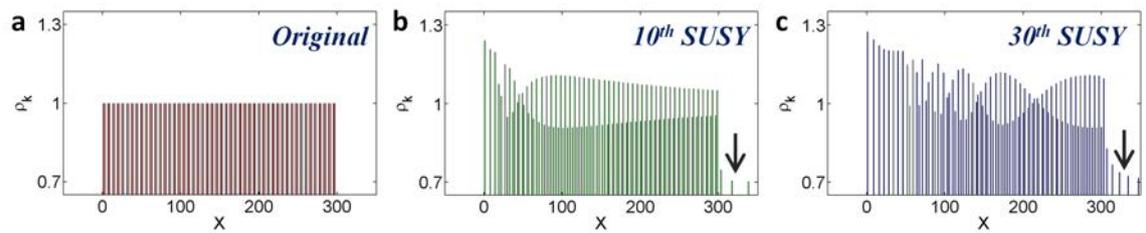

**Supplementary Figure 7. Spatial distribution of optical potentials** for **a,** an original binary photonic molecule, **b,** the 10[th] and **c,** 30[th] SUSY-transformed photonic molecules (corresponding to CMT parameters used in Supplementary Fig. 5). The black arrows denote the positions of decoupled atoms. To convert the coupling coefficients to the physical locations of each atom, the coupling coefficients are defined by setting two conditions: $\kappa = 0.1$ for $\Delta x = 3$ and $\kappa = 0.2$ for $\Delta x = 1$.

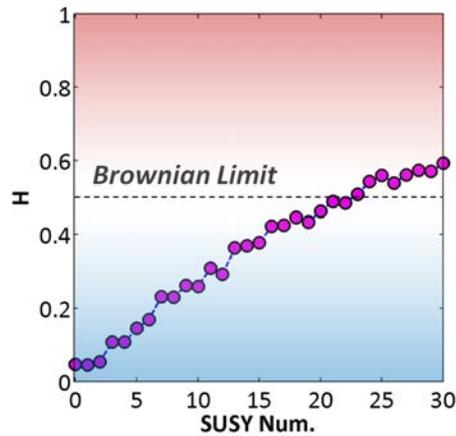

**Supplementary Figure 8. Hurst exponents *H* for each transformed photonic molecule** as a function of the number of SUSY transformations. The red (or blue) region represents the regime of positive (or negative) correlation, whereas the white region corresponds to the uncorrelated Brownian limit.

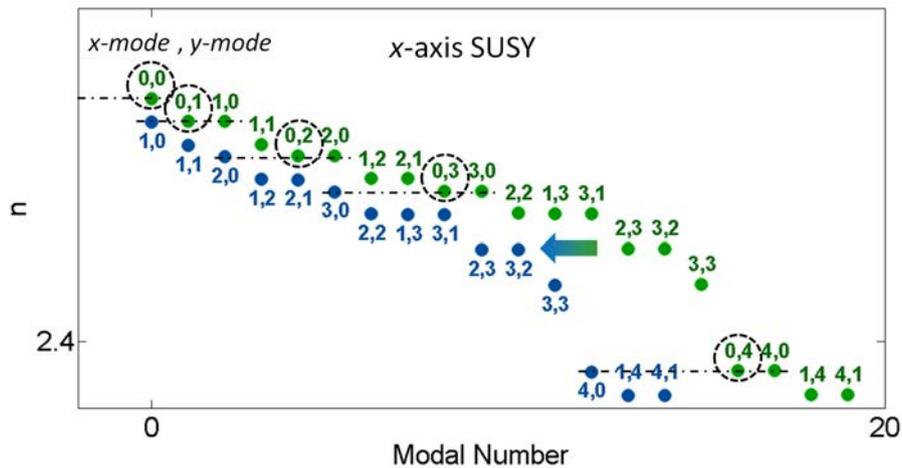

**Supplementary Figure 9. The annihilation of the eigenstates from the *x*-axis SUSY transformation in 2D potentials.** Out of various original eigenstates (green symbols), only the eigenstates with the $0^{th}$ mode profile along the *x*-axis (black dotted circles) are annihilated. The blue symbols are SUSY-transformed eigenstates. The initial shapes of the potentials are the same as those used in Fig. 5 of the main manuscript.

**Supplementary Note 1. The localization based on SUSY transformations**

Supplementary Figure 1 shows examples of SUSY-transformed eigenstates (from the $0^{th}$ to the $3^{rd}$ modes, $M\psi = \{ik / k_0 + \partial_x \psi_0(x) / [k_0 \psi_0(x)]\} \cdot \psi)$ in crystals and quasicrystals. In the case of the crystal with eigenstates that have highly overlapping intensity profiles, the 'bound' profile of $\psi_0(x)$ decreases the spatial bandwidth of each eigenstate with the contribution of $\{\partial_x \psi_0(x) / [k_0 \psi_0(x)]\} \cdot \psi$. However, because the eigenstates in the quasicrystal are already spatially separated, in contrast to those in the crystal, the spatial modification by $\psi_0(x)$ occurs in a much more complex manner (Supplementary Figs 1e-1h).

To quantify the localization of eigenstates in SUSY-transformed potentials, we introduce the definition of the effective width[1] $w_{eff}$ based on the inverse participation ratio, as

$$w_{\text{eff}} = \frac{\left[\int I(x)dx\right]^2}{\int I^2(x)dx},\tag{1}$$

where $I(x)$ is the intensity of each eigenstate. Supplementary Figure 2 shows the change of effective width for the 30 lowest eigenvalue modes by serial SUSY transformations (arrows in Supplementary Fig. 2). As shown, the effective width decreases gradually in the crystal potentials, whereas no tendency is observed in the quasicrystal potentials. However, for both crystals and quasicrystals, the cases of localized eigenstates were found from serial SUSY transformations.

**Supplementary Note 2. Bloch-like waves in discrete optical systems based on coupled mode theory**

The notion of discrete optical systems has played a critical role in the design of optical devices. By interpreting the actual landscape of optical potentials (permittivity and permeability) as a network of optical elements (or 'photonic molecule'[2,3] composed of 'photonic atoms'), the flow of light can be understood through simplified calculations without the need to solve Maxwell's equations directly. Classically, guided-wave platforms have been considered as the network composed of waveguides and resonators[4], and photonic crystals of various symmetries have been treated as tightly bound networks of dielectric atoms[5]. Recently, metamaterials have been investigated in the context of the interplay between electric and magnetic dipoles with elementary oscillations[6,7], *e.g.*, the Lorentz model.

Coupled mode theory (CMT)[4] is a powerful technique for investigating discrete optical systems based on the 1[st]-order approximation of wave equations, not only for the description of spatial beam dynamics in guided-wave platforms[8] but also for the analysis of coupled resonances[9] or metamaterials[10] in the temporal domain. Due to its simplified and generalized formulation with well-matched results, CMT have also been applied to the investigation of Bloch optical potentials, as shown in the studies of waveguide grating[11], Bloch oscillation[12], and the slow light structure based on coupled-resonator optical waveguides (CROW)[13].

Extending the discussion of the main manuscript, in Supplementary Note 2, we derive Bloch-like wave families based on supersymmetry (SUSY) in 'discrete' optical systems by applying CMT to the notion of photonic molecules[2,3] with periodicity.

**Hamiltonian equation of photonic molecules based on CMT**

In a 1-dimensional (or 'nearest-neighbor' coupling) problem, the elementary equation of CMT for the $k^{\text{th}}$ photonic atom is as follows:

$$\frac{d}{d\xi}\psi_k = -i\rho_k\psi_k + i\kappa_{k,k-1}\psi_{k-1} + i\kappa_{k,k+1}\psi_{k+1}, \qquad (2)$$

where $\xi$ is a spatial axis $x$ (or a time axis $t$) in the spatial (or temporal) CMT, $\psi_k$ and $\rho_k$ are the field amplitude and the self-evolution term of the $k^{\text{th}}$ atom, respectively ($\rho_k$ is a wavevector in a spatial CMT or a resonant frequency in a temporal CMT), and $\kappa$ is the coupling coefficient between atoms. The governing equation of a photonic molecule, composed of $N$ photonic atoms, can then be expressed as the eigenvalue equation $H_0\Psi_e = \rho_e\Psi_e$ where $\rho_e$ is an eigenvalue, $\Psi_e$ is a corresponding eigenvector, and $H_0$ is a Hamiltonian matrix in a tridiagonal form,

$$H_0 = \begin{bmatrix} \rho_1 & -\kappa_{12} & & & \\ -\kappa_{21} & \rho_2 & -\kappa_{23} & & \\ & -\kappa_{32} & \rho_3 & \ddots & \\ & & \ddots & \ddots & -\kappa_{N-1,N} \\ & & & -\kappa_{N,N-1} & \rho_N \end{bmatrix}. \qquad (3)$$

If the photonic molecule is Hermitian without magneto-optical effects, the coupling between photonic atoms is symmetric[4] as $\kappa_{ij} = \kappa_{ji}$.

**Band analysis of crystalline photonic molecules**

Consider the case of 1-dimensional crystalline photonic molecules satisfying the Hermiticity ($\kappa_{ij} = \kappa_{ji}$, $\kappa_{k,k+1} = \kappa_k$). Here, we investigate the binary atomic distribution in the same way as in the main manuscript (CMT parameters of identical $\rho_k = \rho_0$ for all $k$, and $\kappa_{2m+1} = \kappa_a$, $\kappa_{2m+2} = \kappa_b$ for $m = 1, 2, \ldots$). Supplementary Figure 3 shows the schematics of binary photonic molecules, each for the waveguide-based example with a spatial CMT model[8] and the resonator-based example with a temporal CMT model[9].

Supplementary Figure 4 shows CMT-calculated eigenspectra as a function of modal number for the discrete system shown in Supplementary Fig. 3. Due to the binary arrangement, a bandgap is achieved around the self-evolution term $\rho_0 = 1$. Note that the construction of the bandgap can be understood in terms of the repulsion (determined by $\kappa_a$) of even and odd parity-modal bands (formed by $\kappa_b$)[14]. By increasing the number of photonic atoms $N$, the eigenspectrum approaches the continuous band with a well-known cosine form[13,14], which is the feature of the lattice having *periodic* couplings (Supplementary Fig. 4a). The width of the bandgap is determined by the contrast between $\kappa_a$ and $\kappa_b$, which is the same as other photonic crystals (Supplementary Fig. 4b).

**SUSY-based random-walk photonic molecules: CMT modelling**

To achieve the SUSY transformation for the CMT-based Hamiltonian equation $H_o \Psi_e = \rho_e \Psi_e$ with the process of the ground-state ($\rho_{e0}$) annihilation, the modified matrix $H_o' = H_o - \rho_{e0} I$ should be decomposed as $H_o' = M^\dagger M$, which is identical to the main manuscript. Due to the Hermitian and positive-definite features of $H_o'$, we apply the Cholesky decomposition as demonstrated in the phase-matching design based on SUSY transformations (refs 15-17), which gives the upper triangular matrix $M$. The SUSY-transformed Hamiltonian $H_s$ is then defined as $H_s = MM^\dagger + \rho_{e0} I$. Note that because $H_o'$ is a tridiagonal matrix, $MM^\dagger$ and thus $H_s$ also become tridiagonal matrices. Therefore, the SUSY-transformed CMT models with $H_s$ maintain the dimension of an original CMT model, allowing only nearest-neighbor couplings[15-17]. Supplementary Figure 5 shows the variation of CMT parameters after the series of SUSY transformations of the binary photonic molecule in Supplementary Fig. 3.

In accordance with the varying amplitudes and frequencies of the potential shapes in the SUSY-transformed continuous potentials (in the main manuscript), the distributions of self-evolutions (Supplementary Fig. 5a) and interatomic couplings (Supplementary Fig. 5b) both become disordered after SUSY transformations. Note that ground-state annihilations are expressed in the form of the decoupling (black arrows)[15-17]. Decoupled atoms thus have self-evolution value lower than $\rho_0$ (Supplementary Fig. 5a).

Supplementary Figure 6 shows the eigenspectrum of each SUSY-transformed photonic molecule. Identical to the case of continuous potentials in the main manuscript, all of the spectral information in the original eigenspectrum (Supplementary Fig. 6a) are preserved during the series of SUSY transformations (Supplementary Figs 6b,6c), including the width and position of the bandgap and eigenbands. Note that the well-known cosine form of the eigenbands, which had been believed to originate from the *periodic* coupling[13], is also reproduced perfectly with disordered photonic molecules. Due to the ground-state annihilation, the lowest part of the SUSY-transformed eigenspectrum has eigenvectors localized to decoupled atoms (black arrows).

**SUSY-based random-walk photonic molecules: real space design**

To design the real structure corresponding to the CMT parameters used in Supplementary Fig. 5, the position and self-evolution term (a wavevector in a spatial CMT, and a resonant frequency in a temporal CMT) of each photonic atom should be determined. While the self-evolution can be easily manipulated through the design of photonic atoms, the coupling is mainly determined by the interatomic distance. The coupling coefficient is generally obtained as[4]

$$\kappa_{ij} = -\frac{i\omega}{4} \int \Delta\varepsilon \cdot \hat{e}_i \cdot \hat{e}_j^* \, ds \, , \tag{4}$$

where $\Delta\varepsilon$ is the perturbation of permittivity distribution and $e_k$ is the normalized field pattern of the $k^{th}$ photonic atom. In the weak coupling regime, based on the evanescent field overlap, the coupling coefficient can be approximated as $\kappa_{ij} \sim c_1 \cdot exp(-c_2 \cdot \Delta x_{ij})$ where $c_{1,2}$ are platform-dependent constants and $\Delta x_{ij}$ is the distance between the $i^{th}$ and $j^{th}$ photonic atoms. Two unknown constants $c_{1,2}$ are determined when $\kappa_{ij}$ for two different distances are defined, and then from $c_{1,2}$, all of the coupling coefficients in Supplementary Fig. 5 can be converted to actual physical locations. The spatial distributions of the photonic molecules (obtained from Supplementary Fig. 5) are shown in Supplementary Fig. 7, presenting the spatially disordered potential shape after the SUSY transformations.

From the results in Supplementary Fig. 7, we can now calculate the correlation of the potential shapes by applying the Hurst exponent[18,19]. Supplementary Figure 8 shows the Hurst exponents for the SUSY-transformed binary photonic molecule, as a function of the number of SUSY transformations. In agreement with the results in the main manuscript, although the original binary molecule has a strong negative correlation, the degree of long-range disorder in SUSY-transformed molecules increases rapidly with the series of SUSY transformations. The disorder comparable to the Brownian limit is also achieved, exhibiting a transition between negative and positive correlations.

In this Supplementary Note 2, we demonstrated the design of SUSY-based Bloch-like potentials in CMT-modelled discrete optical systems, perfectly preserving the width and position of bandgaps and the shape of each eigenbands. By employing the CMT implementation from the $1^{st}$-order approximation of Maxwell's equations, we showed that SUSY randomization of the potential can be applied to the system composed of generalized optical elements, transparent to polarizations and forms of eigenstates. Following this approach, the design of spatial or temporal Bloch-like wave devices with tunable correlations should be possible. For example, slow light propagation along disordered structures by SUSY-transforming the CROW, while preserving all of the spectral information, such as group velocity and its dispersion, can be envisaged. With the polarization-transparency of the CMT, the design of polarization-independent bandgaps can be expected as well, using dual-polarized optical elements[20]. The matrix-based SUSY randomization can also be extended into other basis systems allowing discretization, such as tight-binding analysis, plane-wave expansion methods, and density functional theory in quantum mechanics.

**Supplementary Note 3. The annihilation of eigenstates from 2D SUSY transformations**

Starting from the potential with the form $V_o(x,y) = V_{ox}(x) + V_{oy}(y)$, it can be shown that the eigenstates of the 2D potential are combinations of the eigenstates from the 1D potentials $V_{ox}(x)$ and $V_{oy}(y)$. By using the separation of variables for the 2D Schrodinger-like equation with an eigenstate $\psi(x,y) = \varphi(x) \cdot \phi(y)$,

$$\left[ -\frac{1}{k_0^2} \cdot \frac{1}{\varphi} \cdot \frac{d^2\varphi(x)}{dx^2} + V_{ox}(x) \right] + \left[ -\frac{1}{k_0^2} \cdot \frac{1}{\phi} \cdot \frac{d^2\phi(y)}{dy^2} + V_{oy}(y) \right] = \gamma, \qquad (5)$$

each brace should be a constant, which is one of the eigenvalues of the 1D Schrodinger-like equation, with the potential $V_{ox}(x)$ or $V_{oy}(y)$ (green symbols in Supplementary Fig. 9).

Following the discussion in the Methods in the main manuscript, the annihilation by 2D SUSY transformations occurs not only in the ground state but also in all of the excited states that share a common 1D ground-state profile. The example of this phenomenon is shown in Supplementary Fig. 9 for the *x*-axis SUSY transformation.

**Supplementary References**